\documentclass[referee]{aa} 
\usepackage{graphicx}
\usepackage{txfonts}
\usepackage{natbib}
\begin{document}
   \title{Gas Phase diagnostics of Protoplanetary disk extension}

\author{B.~Ercolano\inst{1}, J.J.~Drake\inst{2}, C.J.~Clarke\inst{1}}

\institute{Institute of Astronomy, University of Cambridge, Madingley Road, 
Cambridge, CB3 0HA,  UK\\
\email{be@ast.cam.ac.uk; cclarke@ast.cam.ac.uk}
\and
Smithsonian Astrophysical Observatory, MS-3, 60 Garden Street, 
Cambridge MA~02138\\
\email{jdrake@cfa.harvard.edu}
}

   \date{Received; accepted }

 
  \abstract
{}
   {We investigate the potential of using ratios of fine structure and
     near-infrared forbidden line transitions of atomic carbon to diagnose
     protoplanetary disk extension.} 
{Using results from 2D photoionisation and radiative transfer modeling of a 
realistic protoplanetary disk structure irradiated by X-rays from a
T~Tauri star, we obtain theoretical emission maps from which we
construct radial distributions of the strongest emission lines produced in the disk. }
{We show that ratios of fine structure to near-infrared forbidden line emission of atomic carbon are especially promising to constrain the minimum size of gaseous protoplanetary disks. While theoretically viable, the method presents a number of observational difficulties that are also discussed here. }
   {}


\titlerunning{Gas Phase diagnostics of Protoplanetary disk extension}
\authorrunning{B.~Ercolano et al.}
\maketitle


\section{Introduction}

Characterizing the gas distribution of protoplanetary disks is
fundamental to many aspects of star and planet
formation, as disks provide the reservoir of material for nascent
stellar-planetary systems. In particular, the outer radius is a key
parameter for understanding a disk's viscous evolution and its
ability to form planets. Indeed, it is the viscous timescale at the
outer edge of a disk that controls the rate at which the 
flow of gas and dust into the inner disk declines (e.g. Hartmann et
al. 1998), and it is therefore relevant to planet formation at small
radii.  

The sizes of protoplanetary disks have been determined from direct
imaging only in a few cases, e.g., in the optical for sources in the
Orion Nebula Cluster (ONC), where the disk silhouettes can clearly be
distinguished (e.g.\ O'Dell 1993, Bally et al.\ 1998) and in the millimeter and
submillimeter for some nearby objects (e.g., Kitamura et al. 2002;
Andrews \& Williams 2008; Wilner et al. 2003; 
Lommen et al. 2007; Testi \& Leurini 2008, and references therein).
Observations show evidence that protoplanetary disks are
photoevaporated to smaller sizes by the intense UV flux in cluster
environments. Rodmann (2002) analysed the silhouette disks in the ONC
and found that 80\% of systems hosting 
proplyds\footnote{Identified as compact ionised gas around low mass
  stars. These systems often show offset cometary-shaped ionisation fronts indicating photoevaporation} contained disks unresolved by the {\it Hubble
  Space Telescope}, and hence smaller than 50~AU. This contrasts with
the submillimeter observations of Kitamura et al. (2002) and Andrews
\& Williams (2008) in the Taurus-Auriga and Ophiuchus-Scorpius star
forming regions, implying a strong environmental dependence. 

 Models of sub-solar and solar-mass systems
(e.g. Fatuzzo \& Adams, 2008; Clarke 2007; Adams et al., 2004; Hollenbach et
al. 2000) predict that in a cluster environment 
protoplanetary disks are photoevaporated from the outside in by a
strong EUV and FUV radiation field.  This has obvious implications for
planet formation, which is generally considered to be compromised when
the disk evaporation time becomes less than $\sim$10 Myr at a disk
radius of $\sim$30~AU.  Fatuzzo \& Adams (2008) conclude that for the
distribution of cluster size in the solar neighbourhood, 25\%\ of the
disk population lose their planet forming potential due to FUV
radiation and 7\%\ due to EUV radiation.  The effects of
photoevaporation on planet formation are predicted to be even more
dramatic for lower mass systems (e.g. Laughlin et al., 2004). 
 However, such models have yet to be verified by
observations.  Comparison of gaseous disk sizes and ages in a variety
of environments would help constrain the efficiency of the external
photoevaporation process.

For all but the closest systems, however, direct imaging is not
possible and disk sizes can only be inferred using indirect techniques.
A commonly applied indirect method to determine disk sizes
relies on modeling of observations of dust continuum emission.  
However, the interpretation of the continuum spectral
energy distribution (SED) relies on dust radiative transfer modeling
and is complicated by a number of known degeneracies between different
model parameters.  For example, the radial distance of the emitting
dust (i.e. the size of the dust disk) is degenerate with respect to
the temperature of the irradiating source (stellar,
interstellar/cluster radiation fields), with respect to the assumed
grain size (which is highly uncertain in disks due to grain growth),
and (less dramatically) to the disk chemical state.  Further, the
temperature of the grains is sensitive to the geometry of the disk
(flaring) and to dust settling.  These aspects are poorly constrained
and lead to significant model-dependence in the SED interpretation
(e.g. Alcal\'a et al. 2008).
A further problem with the determination of disk sizes from dust SED
modeling is that cold dust at large radii only has a very small effect
on the emergent SED and may remain undetected.  Finally, the estimates
of disk size will also depend on whether the disk models used assume a
smooth dust distribution or a clumpy one (e.g., Ercolano et al.,
2007).

Regardless of the accuracy of the disk size calculations from SED
dust modeling, one should also consider circumstances in which the
extension of the dust disk is not a good tracer of the real gaseous
disk extension. Throop \& Bally (2005), for example, predict a scenario
where the gas and the small grains in the outer disks are
photoevaporated, while larger grains grow and settle. This would lead
an SED analysis to infer larger outer radii than the real gaseous disk
extension. 
 The opposite scenario has also been suggested based on recent
 observations of the extent of dust continuum emission whereby 
the outer disk appears {\em depleted} in dust, perhaps through inward 
migration (e.g. Isella et al. 2007), while Hughes et al. (2008) show
that employing different disk models can resolve the apparent
discrepancy. The ongoing debate in the literature underlines the
difficulty in obtaining accurate measures of disk size from current
observations. 

While CO ground-based observations have been employed to obtain estimates
of disk sizes (e.g. Dent et al. 2005), gas-phase atomic line
diagnostics have so far yet to be used to sample 
gaseous protoplanetary disk sizes. In fact observations of emission
lines from gaseous disks have only recently become possible in the IR
from space with Spitzer (e.g Lahuis et al. 2007, Espaillat et al. 2007,
Pascucci et al. 2007) and from the ground with SMA 
(Qi et al. 2006) and MICHELLE at Gemini North (Herczeg et al. 2007). 
 Future instrumentation (e.g. Herschel,
ALMA, CARMA, JWST) hold the promise for better spectral coverage,
resolution and signal to noise (e.g. van Disheock \&
Jorgensen 2008).  Theoretical identification of viable
gas phase diagnostics in support of future observing campaigns is
therefore timely.  In this paper we suggest that fine
structure to near infrared forbidden line emission ratios of atomic
carbon might provide a useful measure of the {\it minimum} size of a
gaseous T~Tauri disk.

\section{Atomic carbon diagnostics of disk extension}
 
The ratio of fine structure (FS) to near-infrared (NIR) lines of
neutral carbon increases as the disk radius gets larger. The main
reason for this is the exponential dependence on temperature of these
collisionally excited lines that dominates the emitted flux of the NIR
lines. The NIR lines are approximately 15,000~K (C~{\sc i} 9826~{\AA} 
and 9852~{\AA}) and 31000~K (C~{\sc i} 8729~{\AA}) above ground, which
makes their emissivity negligibly small at the cool temperatures of
the outer disk. FS lines, on the other hand, are only 20-40~K above the
ground and therefore they are not very temperature-sensitive. 

 A comparison of C~{\sc i} FS and NIR lines of different excitation
 temperatures can therefore, at least in
principle, provide a diagnostic for the extension of disk gas. 
The ratio of these lines is to be preferred to using the absolute flux
of C~{\sc i} FS as an indicator as the latter requires an assumption of carbon
depletion onto grains, which may be different in different
systems. This problem would indeed affect the estimates of disk extension
provided by CO measurements.

We have used the thermal and ionisation structure for a typical
T~Tauri disk modeled by Ercolano et al. (2008a, Paper~I) using the MOCASSIN
code (Ercolano et al. 2003, 2005, 2008b) to calculate
spatially-resolved emissivities of transitions of neutral carbon.  
We assume
that the direct  radiative input to the gas is from X-ray irradiation;
at greater depths in the disc, gas-grain collisons are important. This
latter thus incorporates one aspect also of heating by FUV radiation,
since we adopt the grain temperature distribution computed by d'Alessio et al. (2001), which includes the effects of irradiation by the star's optical/FUV
spectrum. We do not include the other main aspects of FUV heating, i.e. 
FUV photoelectric heating, 
$H_2$ vibrational pumping, nor cooling via CO rovibrational lines. FUV
photoelectric heating, in particular, may be an important thermal channel, and its
omission may cause an error in the gas temperatures derived. This effect
will be included in future models. 

While we refer to Paper I for a detailed description of the model,
it is important to give here at least a brief overview of the
gas thermal balance. In the region of
the disc producing the lines discussed here (i.e. at relatively low
column densities), the main heat input is from X-ray
photoionisation, while cooling occurs via
collisionally excited lines of metals and Ly$\alpha$, free-free and free-bound
continua and dust-gas collisions. Figure~5 of Paper~I shows the
relative heating and cooling rates as a function of vertical column at
a radial distance of 1~AU for the model in use. 

We used a fixed density distribution for the the disk, which was
calculated by d'Alessio et al. (2001, 2003, 2005) to fit the median SED of
T~Tauri stars in Taurus. We refer the reader to D'Alessio et al. (2001) for a
detailed description of the model ingredients and calculation.  In
brief, the input parameters for this model include a central star of
0.7~M$_{\odot}$, 2.5~R$_{\odot}$ irradiating the disk with an
effective temperature of 4000~K. Additional disk parameters consist of
a mass accretion rate of 1.e-8 M$_{\odot}$ yr$^{-1}$ and a viscosity
parameter $\alpha$~=~0.01. The total mass of the disk is  0.027~M$_{\odot}$.

We have recalculated the emitted flux of the FS lines adding
 excitation by impact with atomic hydrogen, which is important in the
 outer disk regions and was not treated in Paper~I. Furthermore,
 the line luminosities reported in 
 Table~1 of Paper~I were obtained integrating the disk out to a radius
 of 190~AU. We report here (Table~1) the revised C~{\sc i} FS
 luminosities, which include collisions with atomic H and are
 integrated out to a disk radius of 500~AU. When integrated to a
 radius of 190~AU our revised C~{\sc i} FS line luminosities are
 1.8$\cdot$10$^{-7}$L$_{\odot}$ and 1.1$\cdot$10$^{-6}$L$_{\odot}$ for
     C~{\sc i}~609$\mu$m and C~{\sc i}~370$\mu$m, respectively (a
     factor of 1.4 and 2.8 larger than reported in Paper~I).
The absolute luminosities of the FS lines reported in Table~1 are in
agreement with  Hollenbach et al. (1991) who predicted
the ubiquitous presence of C~{\sc i} emission from cloud surfaces at the
10$^{-5}$erg cm$^{-2}$ s$^{-1}$ level, which corresponds to a luminosity of
 $\sim$2$\cdot$10$^{-6}$~L$_{\odot}$ for a 500~AU disk. 

Excitation by impact with atomic H may also contribute to the emissivities of
the NIR lines in a similar fashion as for the FS lines discussed
above. To our knowledge, the relevant excitation coefficients
are not available in the literature and we are therefore unable to
provide a robust estimate for the magnitude of this effect. The C~{\sc i}
NIR line luminosities reported in Table~1 are thus likely to be
underestimated by a factor of order unity.

\begin{table}
\caption[]{Absolute luminosities of C~{\sc i} collisionally excited
  fine structure and near-infrared lines. FS line luminosities also
  include a contribution from excitation by impact with atomic
  hydrogen.}
\label{t:t1}
\begin{tabular}{cc}
\hline
 Wavelength  & Luminosity   \\
             & $[10^{-7} $L$_{\odot}]$ \\ 
\hline
  8729~{\AA} & 4.6  \\    
  9826~{\AA} & 3.2  \\          
  9852~{\AA} & 9.5  \\          
  370~$\mu$m & 55  \\
  609~$\mu$m & 11  \\
\hline
\end{tabular}
\end{table}

From our theoretical emission maps, we constructed radial distributions of the
strongest emission lines produced in the disk. Figure~1 (right) shows
the normalised line fluxes integrated over the volume of the disk
as a function of disk size. Figure~1 (left) shows the
C~{\sc i} FS to NIR line ratios integrated over the volume of the disk
as a function of disk size (i.e. integration radius).  In our models, values
approaching unity indicate a disk size $\ga$100~AU. The FS/NIR line
ratios for disk sizes of 30~AU are at least a factor of ten smaller   
than those with sizes of 100~AU or larger.  An observational census of 
disks in cluster environments using these line ratios would provide 
a direct observational test of disk photoevaporation predictions
such as those of Fatuzzo \& Adams (2008).

At small radii ($\la$ 30~AU), X-rays produced by the YSO are most
likely the dominant ionising radiation source (e.g. Meijerink et al. 2008,
Glassgold et al. 2004). Although significant amounts of EUV radiation
are emitted by YSOs (Alexander, Clarke \& Pringle, 2005, 
Herczeg \& Hillenbrand, 2008), which may play an important role in
photoevaporating the disc (Shang et al. 2002), it is likely that
such radiation will only be able to penetrate to the disc surface at
late times, when partially ionised winds, launched close to the star,
have become optically thin to Lyman continuum photons. We thus assume
that over the majority of a star's disc bearing lifetime, the EUV
field that can reach the disc at larger radii and depths is
significantly attenuated (Glassgold et al. 2007). External UV radiation
fields, however, may become important at larger radii and may affect
low-excitation lines (see discussion in Section~3).
The models of Ercolano et al. (2008a) did not include 
external UV and FUV radiation fields, and therefore one should further
note this as a potential cause of errors in the calculated neutral carbon column. 
 
   \begin{figure*}  
   \begin{center}
   \includegraphics[width = 0.49\textwidth]{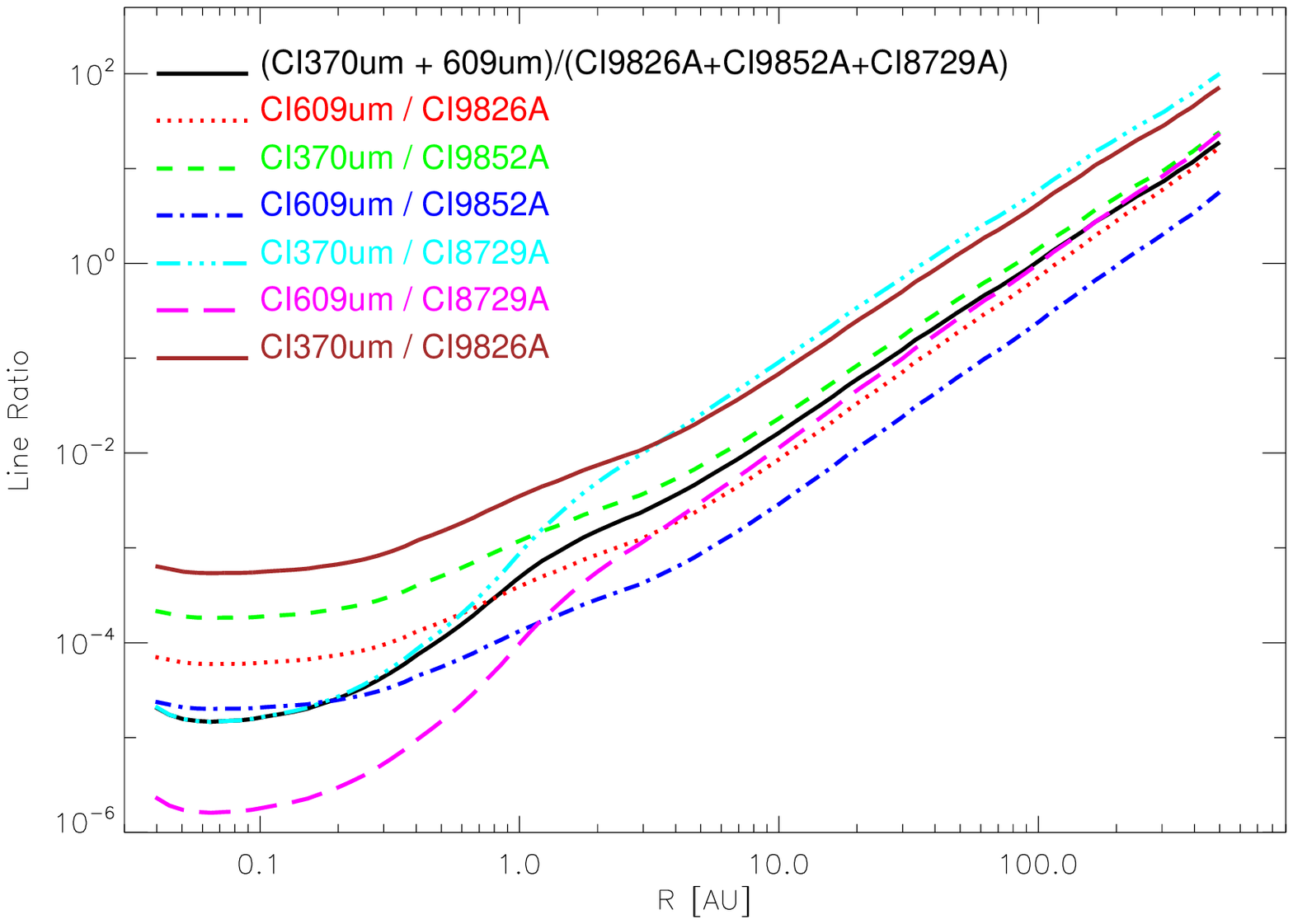}
   \includegraphics[width = 0.49\textwidth]{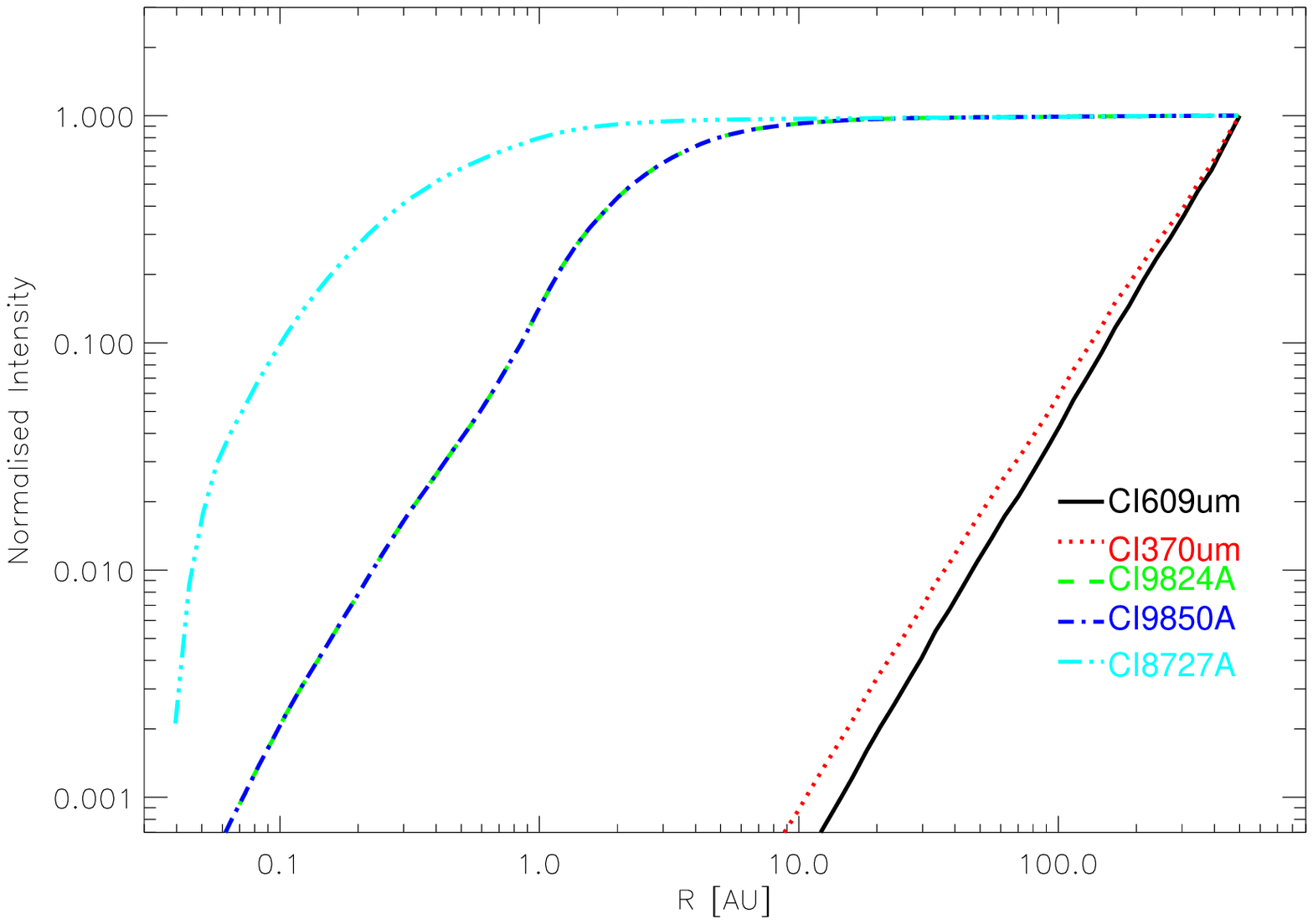}
      \caption{{\it Left:} C~{\sc i} fine structure to NIR collisionally
        excited line ratios integrated over the volume of the disk as
        a function of disk outer radius. {\it Right:} Normalised line
        fluxes integrated over the volume of the disk as
        a function of disk size. Note that [C~{\sc
            i}]~9824.13{\AA} and 9850.26{\AA} have a fixed theoretical
        ratio of 0.34.}
   \end{center}
   \end{figure*}

It should also be mentioned that, as well as collisional excitation of
the ground-state C($^3$P), CO dissociation, 
dissociative recombination of CO$^+$ and collisional dissociation of
CO could all be contributors to the formation of of the C~{\sc i} NIR
lines. In the case of protoplanetary disks, however, we expect collisional
excitation of C($^3$P) to be by far the dominant mechanism for the
emission of the NIR lines, since they are emitted
preferentially in a layer where CO has already
completely dissociated and carbon is mostly atomic or singly
ionised. At 1~AU Glassgold et al. (2004) calculate that the transition from
atomic carbon to CO occurs at a vertical column density of approximately
10$^{21}cm^{-2}$, and, as shown in Figure~2, the bulk of the emitted
flux comes from warm gas at columns of $\sim$10$^{20}$. The electron
temperature in this region, shown by asterisks in Figure~2, is
approximately 4000~K and the electron densities in the NIR emitting
regions are $\sim$10$^4$-10$^5 cm^{-3}$. 
We also note that the lines 9826 and 9852 are 14,668 K above ground and 
the 8729 A is 31154 K above ground, so the latter should be preferentially
 produced in the hotter gas closer to the surface, however higher in the disk
atmosphere carbon becomes ionised causing the emissivity to drop.

   \begin{figure}  
   \begin{center}
     \includegraphics[width = 0.49\textwidth]{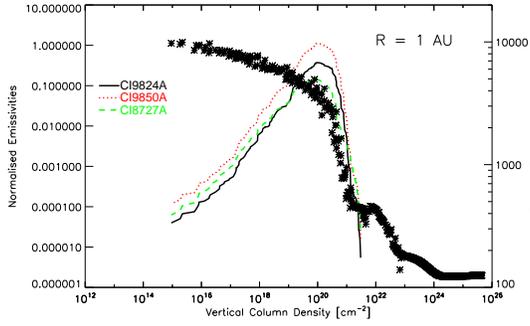}
      \caption{Normalised emissivities of the near-infrared C~{\sc i}
        forbidden lines (solid lines) plotted as a function of
        vertical column density at a radial distance of 1~AU. Electron
        temperatures are also plotted here with black asterisks.}
   \end{center}
   \end{figure}

\section{Discussion}

The gas phase diagnostics we have identified and plotted in Figure~1
offer a promising new avenue for the study of gaseous
T~Tauri disks, but a number of observational considerations should be
taken into account and the detection and measurement of the necessary
emission lines still presents a challenge for current space-based and
ground-based instrumentation.

The main problem facing the ground state forbidden transitions near
1$\mu$m is their appearance against the background photospheric
continuum of the parent T~Tauri star.  We have estimated the expected
line-to-continuum contrast using the emergent photospheric flux
predicted by an {\sc ATLAS12} solar metallicity model atmosphere
(Castelli \& Kurucz 2004; Kurucz 2005), corresponding to the stellar
parameters adopted for the disk models of Ercolano et al.(2008a):
effective temperature $T_{eff}=4000$~K, surface gravity $\log g=3.5$,
and radius $R=2.5R_\odot$.  This model yields equivalent widths for
the [C~{\sc i}] lines in emission of 1.7, 5.0 and 2.2~m\AA\ for
9824.13{\AA}, 9850.26{\AA} and 8727.13{\AA}, respectively.  
While such equivalent widths can in principle be observed, 
these [C~{\sc i}] lines will also be present in absorption in T~Tauri
photospheric spectra.  

These [C~{\sc i}] lines were first investigated in detail in the
solar spectrum by Lambert \& Swings (1967), who found the cleanest
line to be [C~{\sc i}]~8727~\AA, with an equivalent width of 6.5~m\AA.
This line lies in the far wing of a much stronger Si~{\sc i} line, but
is otherwise only blended with a very weak Fe~{\sc i} line whose
contribution to the 8727.13~\AA\ feature is estimated to be a few
percent at most (Lambert \& Swings 1967).  Gustafsson et al.\ (1999)
estimated an upper limit of 15\%\ for the coolest ($T_{eff}\sim
5700$~K), most Fe-rich solar-type star of their study of C abundances
in the Galactic disk.  Instead, the 9824.13{\AA} and 9850.26{\AA}
features were not detected and, while they have been observed in
emission in spectra of planetary nebulae, H~{\sc ii} and
photodissociation regions (e.g.\ Liu et al.\ 1995), they have received
scant attention in the stellar literature in the intervening years.

One exception is the study of R~Coronae Borealis stars by Pandey et
al.\ (2004), who report measurements of [C~{\sc i}]~9850.26{\AA}
obtained close to a neighbouring line of Fe~{\sc ii}, but note that [C~{\sc
  i}]~9824.13{\AA} is irretrievably blended.  Indeed, the latter falls
between telluric water vapour features (which are not necessarily
problematic provided accurate correction can be achieved) but also
between, and on the far wings of, lines of CN.  These CN lines are
very weak in the solar spectrum, with equivalent widths of only a few
m\AA, but are likely to become stronger in the cooler atmospheres of
T~Tauri stars.  We note that a very weak $\sim 1$~m\AA\ feature in the
solar spectrum blueward of [C~{\sc i}]~9850.26{\AA} near 9850.1{\AA},
and possibly the same feature identified as due to Fe~{\sc ii} by
Pandey et al.\ (2004), is also coincident with two lines of Fe~{\sc i}
in the line database of Kurucz (R.L.~Kurucz, personal communication).
A telluric water vapour
line lies to the red of the predicted [C~{\sc i}]~9850.26{\AA}
location at 9850.5{\AA}, but no other conspicuous absorption features
are present.  

In summary, the lack of any direct coincidence between the predicted
location of [C~{\sc i}]~9824.13,9850.26{\AA} and other strong
photospheric lines leads us to conclude that further investigation of
these features, as well [C~{\sc i}]~8727.13{\AA}, for possible
emission signatures in T~Tauri stars is well-motivated.  Emission line
equivalent widths of 1~m\AA\ are clearly challenging in any spectral
range when superimposed on a background of possible blending lines.
For a resolving power of $\lambda/\Delta\lambda \sim 100000$, for
example, 1~m\AA\ corresponds to a peak flux of order 1\%\ of the
continuum at 9000\AA, and requires a signal-to-noise ratio of a few
hundred for a firm detection.  As we discuss below, the uncertainties
inherent in our calculations render our line strength estimates
somewhat uncertain: equivalent widths a few times stronger (or weaker)
are quite possible.  The study of Pandey et al. (2004) using the
McDonald 2.7m telescope and cross-dispersed echelle observed at
$\lambda/\Delta\lambda=120000$ and reached to $J=9$ with
signal-to-noise ratio of $> 100$.  Relatively nearby
T~Tauri stars such as those in the Taurus-Auriga star-forming regions
have typical $J$-band magnitudes of order 9-12. For these ranges of
near infrared magnitudes, equivalent widths of a few m\AA\ are quite
accessible to current high resolution spectrographs on larger
telescopes able to operate in the near-IR.

With regards to the FS C~{\sc i} lines, it should be mentioned here
that both the 370$\mu$m and the 609$\mu$m lines fall in the wavelength
range of 
$ALMA$ and $Herschel$, with good prospect of these lines being
observed in protoplanetary disks in the future. The importance of
these lines in the study of protoplanetary disks was indeed already
recognised by Jonkheid et al. (2006) who made predictions for the
intensities and line profiles for the three [C~{\sc i}] fine structure
lines using their chemical model of the transitional disk around
HD141569A.  These authors point out that neutral carbon, which is the
dominant form of carbon in large parts of the disk, is also a good
tracer of disk mass.

We conclude with a discussion of the modeling uncertainties that may
affect our predicted C~{\sc i} line fluxes.  Ercolano et al. (2008a)
discuss a 
number of limitations of their models, in particular the fact that the
disk density structure was calculated assuming full thermal coupling
between dust and gas, which they show to be a poor approximation in the
warm disk atmosphere.  Furthermore, line ratios may also vary for
different X-ray luminosities  and may also
be sensitive to additional spectral components (for example, the
inclusion of a strong FUV field would affect the abundance of
neutral carbon, both positively, via the photodissociation of
CO, and negatively, through photoionisation of neutral carbon). 
Finally
the C~{\sc i} NIR line intensities are very sensitive to the 
gas temperature in the dominant emitting zone, and therefore
uncertainties in the model-determined thermal structure will affect
the NIR/FS ratio. Although forthcoming models will aim at remedying
these shortcomings, these important caveats should
be taken into account when considering the line diagnostics proposed
here, which are not intended to be used in order to provide precise
measures of disk 
sizes. An observational study based on comparison of fine structure to NIR
forbidden lines of atomic carbon should be carried out in a
statistical fashion, by comparing the mean ratios measured in multiple
objects belonging to different environments, such as dense clusters
like the ONC or low density regions like Taurus-Auriga. An interesting
comparison would also be between regions of different ages. 

The prospect of observational studies based on emission lines produced
in gaseous protoplanetary disks will be greatly improved when {\it
  ALMA}, {\it CARMA}, {\it Herschel} and later {\it JWST} come
on-line. The work presented here aims at contributing to the
construction of a theoretical framework in aid of future observational
campaigns.

\begin{acknowledgements}

We thank the anonymous referee for a thorough and constructive
assessment of our work. JJD was funded by NASA contract NAS8-39073 to
the {\it Chandra X-ray 
Center} during the course of this research.
The simulations were partially run on the Cosmos 
(SGI altix 4700) supercomputer at DAMTP in Cambridge. Cosmos is a 
UK-CCC facility which is supported by HEFCE and STFC.

\end{acknowledgements}

\end{document}